\documentclass[%
reprint,
showpacs,preprintnumbers,
amsmath,amssymb,
aps,
prb,
notitlepage
]{revtex4-1}

\usepackage{graphicx}
\usepackage[export]{adjustbox}
\usepackage{dcolumn}
\usepackage{bm}
\usepackage{wasysym}
\usepackage[utf8]{inputenc}

\begin{document}
\newlength\fheight \newlength\fwidth 
\pacs{75.75.-c, 61.48.Gh, 75.50.Bb, 75.70.Ak}

\title{Stability and magnetization of free-standing and graphene-embedded iron membranes}
\author{M. R. Thomsen}
\author{S. J. Brun}
\author{T. G. Pedersen}
\affiliation{Department of Physics and Nanotechnology, Aalborg University, DK-9220 Aalborg \O st, Denmark}
\affiliation{Center for Nanostructured Graphene (CNG), DK-9220 Aalborg \O st, Denmark}

\begin{abstract}
Inspired by recent experimental realizations of monolayer Fe membranes in graphene perforations, we perform \textit{ab initio} calculations of Fe monolayers and membranes embedded in graphene in order to assess their structural stability and magnetization. We demonstrate that monolayer Fe has a larger spin magnetization per atom than bulk Fe and that Fe membranes embedded in graphene exhibit spin magnetization comparable to monolayer Fe. We find that free-standing monolayer Fe is structurally more stable in a triangular lattice compared to both square and honeycomb lattices. 
This is contradictory to the experimental observation that the embedded Fe membranes form a square lattice. 
However, we find that embedded Fe membranes in graphene perforations can be more stable in the square lattice configuration compared to the triangular. In addition, we find that the square lattice has a lower edge formation energy, which means that the square Fe lattice may be favored during formation of the membrane.
\end{abstract}

\maketitle

\section{Introduction}

In recent years, there has been a tremendous interest in graphene and its derivatives, owing to their remarkable electronic properties, such as ultra-high mobility of 1.000.000 cm$^2$/Vs at low temperature\cite{wang2013one}. These properties make graphene interesting for electronic and spintronic applications. Carbon-based spintronic devices may have a distinct advantage over many other materials in that carbon has a very low spin-orbit coupling together with an absence of hyperfine interaction in the predominant $^{12}$C isotope. This results in long spin lifetimes\cite{han2014graphene,guimaraes2014controlling,drogeler2014nanosecond}, as well as large spin relaxation lengths, which have been found to be on the order of several microns at room temperature\cite{tombros2007electronic,han2014graphene,guimaraes2014controlling,drogeler2014nanosecond} and make graphene ideal for ballistic spin transport\cite{weser2011electronic}.

Pristine graphene is non-magnetic, but several suggestions on how to give graphene magnetic properties have been put forward. Density functional theory (DFT) calculations have shown that ferromagnetism can be introduced in graphene by e.g.\ semi-hydrogenation\cite{zhou2009ferromagnetism}, adding vacancies\cite{lehtinen2004irradiation,haldar2014fe} or adding adatoms\cite{zanella2008electronic,krasheninnikov2009embedding,santos2010first,he2014atomic,rodriguez2010trapping,haldar2014fe}. Semi-hydrogenating graphene sheets, where one sublattice is fully hydrogenated, while the other is not, leads to a sublattice imbalance, which induces a magnetic moment of 1~$\mu_B$ per unit cell\cite{zhou2009ferromagnetism}. 
Monovacancies in graphene have also been demonstrated to have a magnetic moment between 1.04 $\mu_B$\cite{lehtinen2004irradiation} and 1.48 $\mu_B$\cite{haldar2014fe}. Lehtinen \textit{et al.}\cite{lehtinen2004irradiation} find that the spin-polarized state may be unstable, and find that it can be stabilized by adsorption of two hydrogen atoms in the vacancy, with a resulting magnetic moment of 1.2 $\mu_B$. The spin of a vacancy generally increases with the number of missing carbon atoms, except for the divacancy where the magnetic moment is vanishing\cite{haldar2014fe}. Ferromagnetism can also be induced by transition metal adatoms on graphene or in graphene vacancies. Transition metal adatoms in graphene and single-walled carbon nanotubes were studied by Zanella \textit{et al.}\cite{zanella2008electronic} and Fagan \textit{et al.}\cite{fagan2003ab}, respectively. In particular, they find that the spin moment of Fe adatoms is largely unaffected by the presence of carbon. Zanella \textit{et al.} find that the spin moment of Fe adsorbed on graphene is either 2 or 4 $\mu_B$ depending on the adsorption site, while Fagan \textit{et al.} find that the spin moment of Fe adsorbed on a carbon nanotube is about 3.9 $\mu_B$ independent of adsorption site. 
DFT calculations show that a single Fe adatom on a graphene monovacancy is non-magnetic\cite{krasheninnikov2009embedding,santos2010first,he2014atomic}. However, by adding a Hubbard U term to the GGA functional, Santos \textit{et al.}\cite{santos2010first} showed that this state may, in fact, be magnetic with a spin moment of 1 $\mu_B$, and that the non-magnetic properties predicted by the GGA calculation is a consequence of the limitations of the functional itself. Nevertheless, the spin moment of a single Fe adatom on a graphene monovacancy is strongly decreased compared to free Fe, due to the Fe-C interaction. A single Fe adatom in a graphene divacancy, however, has a spin moment of about 3.2~$\mu_B$ according to Krasheninnikov \textit{et al.}\cite{krasheninnikov2009embedding}, and 3.55~$\mu_B$ according to He \textit{et al.}\cite{he2014atomic} 
The reason for the increased spin is quite obvious; the larger vacancy increases the Fe-C distance and thus decreases the interaction between Fe and C. As the interaction between Fe and C seems to decrease the spin moment of Fe, we expect Fe-C systems to have decreased spins compared to a pure Fe system. Trapping larger Fe clusters in graphene perforations will lead to a larger spin moment, which combined with the electrical properties of graphene, might make this a suitable system for graphene-based spintronics.

Trapping of metal atoms, such as Fe and Mo, in graphene and carbon nanotube vacancies have been achieved experimentally in transmission electron microscopy (TEM)\cite{rodriguez2010trapping,robertson2013dynamics}. Vacancies are created under e-beam irradiation, after which mobile metal atoms on the surface move to the vacancy, where they are trapped. These trapped metals are stable for some time, but detrapping of some of the atoms have been observed over time\cite{rodriguez2010trapping,robertson2013dynamics}, which is thought to occur due to weak bonding, e-beam irradiation or due to high temperature during the experiments. 
Recent experimental results by Zhao \textit{et al.}\cite{zhao2014free} show that monolayer Fe membranes can be grown in graphene perforations. These monolayer membranes both form and collapse under e-beam irradiation in TEM. The Fe is provided via leftover residue from the transfer process, where graphene is transferred from growth substrate to target substrate. Electron energy loss spectroscopy (EELS) and high-angle annular dark-field (HAADF) measurements suggest that the embedded membranes are composed of pure Fe. They find that the embedded Fe membranes form a square lattice with a lattice constant of about 2.65~Å. Through density functional theory (DFT) calculations, Zhao \textit{et al.} find that  monolayer Fe is most stable in a square configuration with a lattice constant of 2.35~Å. They argue that the difference between observed and calculated lattice constant may be a result from straining due to lattice alignment and mismatch between the Fe membrane and graphene.

In this paper, we present a DFT analysis of the structural stability and magnetization of Fe systems in an attempt to obtain a basic understanding of these systems, as well as to explain the experimental results by Zhao \textit{et al.}\cite{zhao2014free}. In particular, we compare the stability of Fe in square and triangular lattice configurations for both monolayer Fe, monolayer Fe carbide and Fe embedded in graphene perforations. We model embedded Fe membranes as a periodic system, effectively giving rise to graphene antidot lattices (GALs), where the antidots are filled with Fe. GALs, which are periodic perforations in an otherwise pristine graphene sheet, can be produced experimentally by, e.g., e-beam lithography on pristine graphene\cite{eroms2009weak,giesbers2012charge}. It is possible that the embedding of iron in graphene perforations can be scaled up to actual Fe filled GALs. GALs have tunable band gaps that depend on geometric factors\cite{pedersen2008graphene,brun2014electronic}, which make them interesting for electronic and optoelectronic applications. It has been shown that a narrow slice of GAL with just a few rows connected to graphene sheets on either side is sufficient to block electron transport in the energy gap of the GAL\cite{pedersen2012transport,thomsen2014dirac}. By omitting antidots in some regions of such a GAL barrier, electrons can be guided through the unpatterned part, giving rise to an electronic waveguide\cite{pedersen2012graphene}, reminiscent of a photonic waveguide in a photonic crystal. Iron-filled GALs could be an ideal platform for spintronics if they can combine the high degree of control over electrons with the magnetic properties of Fe.

\section{Theoretical methods}

Spin-polarized DFT calculations were performed using the Abinit package\cite{gonze2002first, gonze2009abinit, bottin2008large, torrent2008implementation}, which uses a plane-wave basis set to expand the wave function. We have used the Perdew-Burke-Ernzerhof GGA (PBE-GGA) exchange and correlation functional\cite{perdew1996generalized} in all calculations. We use a plane-wave cutoff energy of 435~eV combined with the projector-augmented wave (PAW) method\cite{kresse1999ultrasoft}. It has previously been demonstrated that the PAW method is able to accurately describe magnetism in transition metal systems.\cite{kresse1999ultrasoft,kresse2002comment} 
We use a Fermi smearing of 0.27~eV in order for a $16 \times 16 \times 1$ Monkhorst-Pack $k$-point grid to be adequate. The Fermi smearing has the effect of slightly lowering the magnetic moment as electrons will have a probability to occupy states above the Fermi level. An interlayer spacing of 10~Å was used in all calculations. Full relaxation of all atoms in the unit cells were made for all structures, in addition to relaxation of the unit cell size in the case of free-standing monolayer Fe and iron carbide. Atomic coordinates were optimized until the maximum force on atoms was smaller than 0.05~eV/Å. These parameters have previously been shown to be adequate for modeling transition metal adatoms on graphene vacancies\cite{krasheninnikov2009embedding,lehtinen2004irradiation}.

\section{Free-standing monolayer systems}
\subsection{Monolayer iron}

In order to obtain an understanding of iron membranes embedded in graphene perforations, we first determine the stability of free-standing monolayer iron in different lattice configurations. Then, we calculate the edge formation energy of monolayer iron, in order to obtain an understanding of the formation kinetics of iron membranes. Lastly, we determine the stability of iron membranes embedded in graphene antidots for certain hole sizes.

\begin{figure}[htb]
	\centering
	\includegraphics{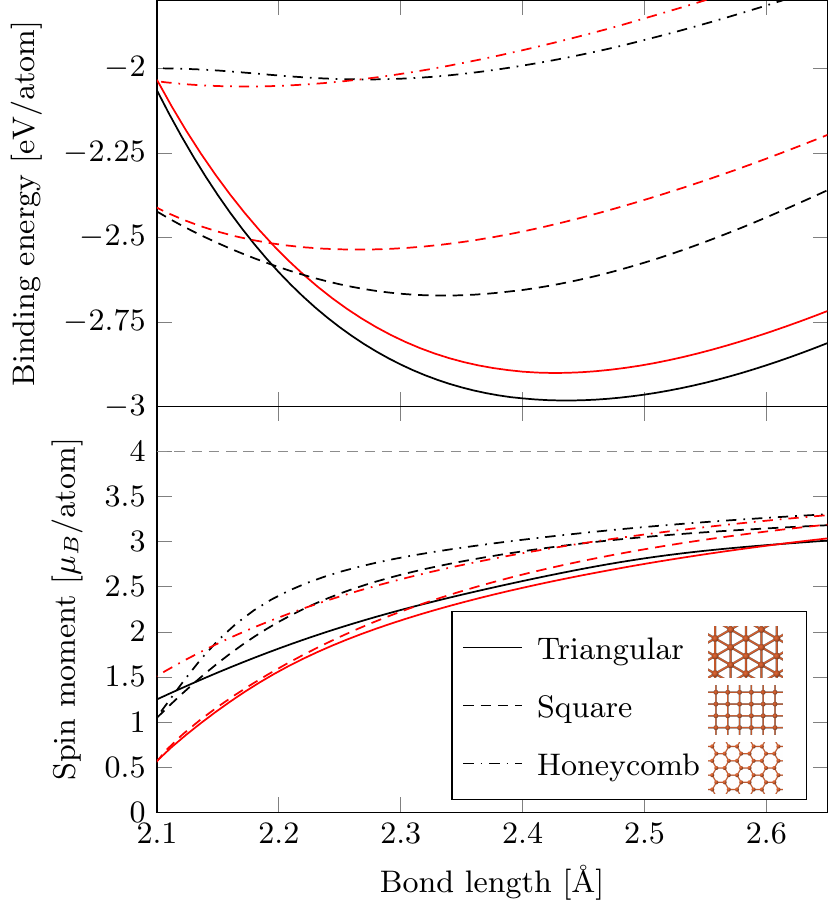}
	\caption{Binding energy (upper panel) and spin moment (lower panel) of monolayer Fe as a function of bond length. The black and red lines are for ferromagnetic and antiferromagnetic ordering, respectively. The magnitude of the spin is shown in case of antiferromagnetic ordering, as it has zero net spin. The dashed gray line indicates the spin of free Fe.}
	\label{fig:Fe_lat_const_binding_magnet}
\end{figure}

The binding energy and magnetization of free-standing monolayer iron in square, triangular and honeycomb lattice configurations are shown in Fig. \ref{fig:Fe_lat_const_binding_magnet}. The figure shows that ferromagnetic ordering is generally favored over antiferromagnetic ordering, consistent with earlier results which shows that monolayer Fe in the square lattice favors ferromagnetic ordering\cite{blugel1989magnetic}. The figure also shows that the honeycomb lattice is unfavored compared to the square and triangular lattices. 
We therefore exclude antiferromagnetic ordering as well as the honeycomb lattice in the remaining calculations. In addition, the figure shows that the most stable configuration is the ferromagnetic triangular lattice, as it has the lowest binding energy at equilibrium. However, it is seen that, under compressive strain, the ferromagnetic square lattice eventually becomes favored. The spin moments per atom at equilibrium are 2.73 $\mu_B$ and 2.68 $\mu_B$ for the square and triangular lattice, respectively, which is significantly larger than the bulk spin moment of 2.22 $\mu_B$\cite{haynes2014crc}. Our results for the spin of the ferromagnetic triangular lattice are in good agreement with previous results.\cite{boettger1993strain,achilli2007ab}

As expected, we see that the spin moment increases with increasing distance between the Fe atoms, as the spin tends towards 4 $\mu_B$ for free Fe. We notice that the bond length at equilibrium of the square lattice is 2.33~Å, which is significantly lower than the experimental results of 2.65~Å by Zhao \textit{et al.}\cite{zhao2014free}, suggesting that the Fe membranes are strained by the surrounding graphene. In addition, it is seen that the energy cost of straining the square lattice to 2.65~Å is only about 0.2 eV per atom. Our predictions of the lattice constant and energy cost of straining for the square monolayer Fe lattice are very close to the theoretical results by Zhao \textit{et al.}. 
The major difference between the results is that we find the triangular lattice to be more stable, whereas Zhao \textit{et al.} find that the square lattice is more stable, in agreement with their experiments. 
Despite the fact that Zhao \textit{et al.} find their theoretical results to be in agreement with experiment, we find them to be inaccurate for two reasons. 
First, Zhao \textit{et al.} use a Monkhorst-Pack $k$-point sampling of only 3$\times$3$\times$1, which we find to be insufficient to describe both spin magnetization and total energy, especially without any temperature smearing. In our calculations, we have carefully tested for convergence by systematically increasing the density of $k$-points. Second,  Zhao \textit{et al.} use a localized basis set, which is much more prone to systematic errors than plane-wave basis sets, as it is difficult to choose additional basis functions to increase accuracy, whereas one can always add more plane-waves to a plane-wave basis to increase accuracy. Therefore, calculations using localized basis sets should always be verified by e.g.\ comparing with results obtained in a plane-wave basis. 
Due to the insufficient $k$-point sampling and possible systematic errors in the basis set, we believe that the accuracy of our results is superior to those by Zhao \textit{et al.}

\subsection{Edge energy of monolayer iron}

\begin{figure}[htb]
	\centering
	\includegraphics{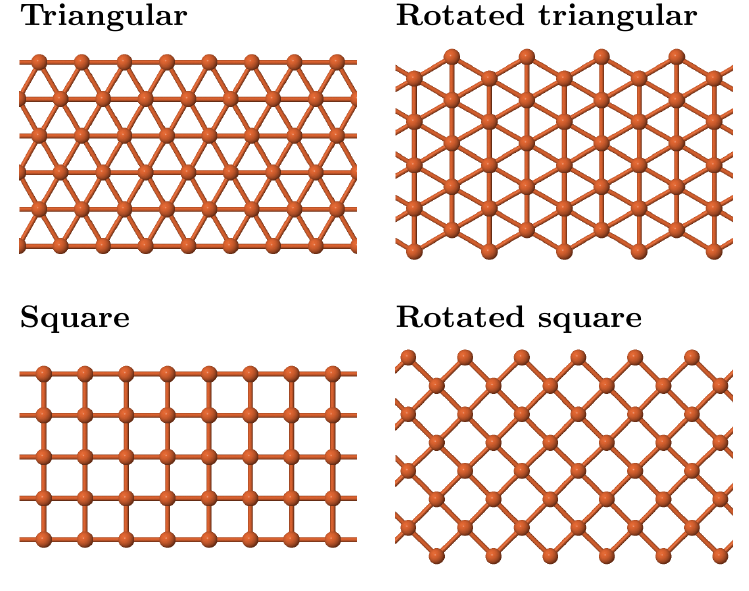}
	\caption{Geometries used for evaluation of edge energies.} 
	\label{fig:edge_structures}
\end{figure}

We have demonstrated that the triangular lattice is energetically favored over the square lattice, so in order to explain why the square lattice is formed experimentally, we now analyze the edge formation energy by comparing the energy of an Fe nanoribbon and monolayer Fe. The edge formation energy per length is given by $E_{edge}=(E_{ribbon}-NE_{monolayer})/2l$, where $l$ is the length of the unit cell in the direction of the ribbon edge, $E_{ribbon}$ is the total energy of the nanoribbon unit cell, $N$ is the number of atoms in the unit cell and $E_{monolayer}$ is the energy per atom of the monolayer system. The factor of $1/2$ is due to the fact that a nanoribbon has two edges. For both the square and the triangular lattice, we examine two different rotations of the edges, as shown in Fig. \ref{fig:edge_structures}.

\begin{figure}[htb]
	\centering
	\includegraphics{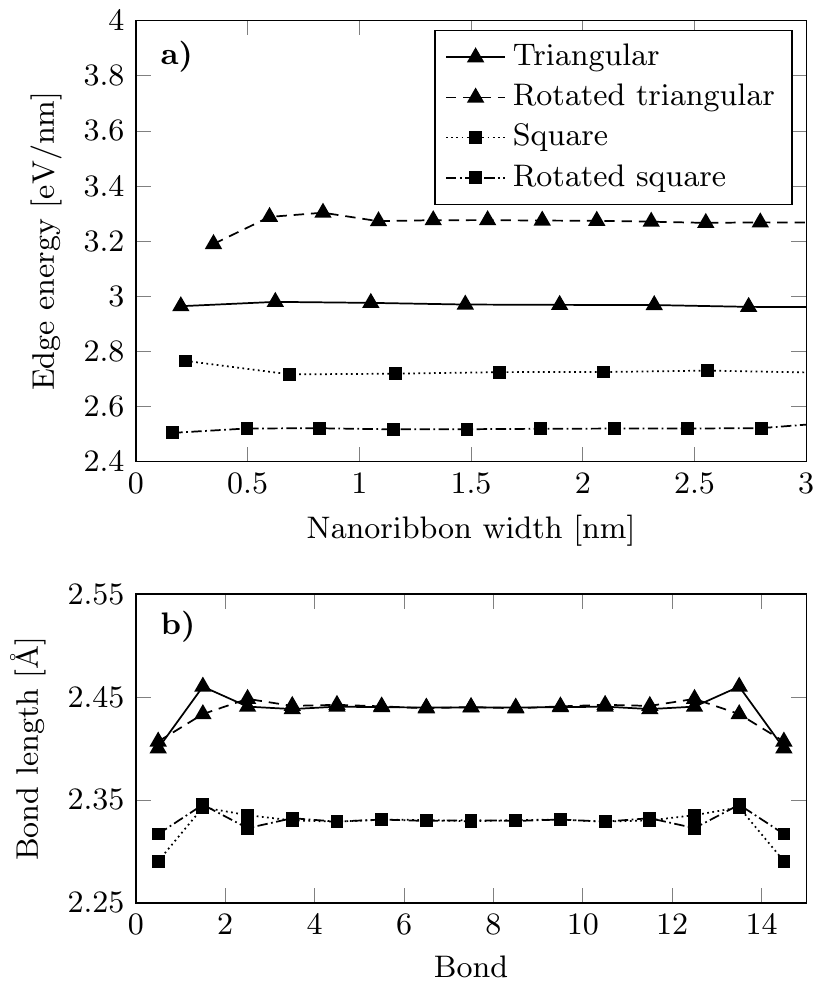}
	\caption{a) Edge formation energy for square and triangular Fe nanoribbons as a function of nanoribbon width. b) Bond lengths through a 16-atom-wide Fe nanoribbons with different orientations and edge rotations.}
	\label{fig:edge_energy_and_bond_length}
\end{figure}

In Fig. \ref{fig:edge_energy_and_bond_length}a we observe that the triangular lattice has a larger edge formation energy than the square lattice for both rotations of both lattices. This means that, during formation of the membrane, the square lattice may be favored due to the lower edge formation energy. The membrane may then be kinetically hindered from subsequently rearranging into the triangular lattice. It is seen in Fig. \ref{fig:edge_energy_and_bond_length}b that the bond length contracts on the edges of the ribbon, while the remaining structure is almost unchanged. This indicates that the large experimentally observed lattice constant is not due to formation kinetics.

\subsection{Iron carbide}
\begin{figure}[htb]
	\centering
	\includegraphics{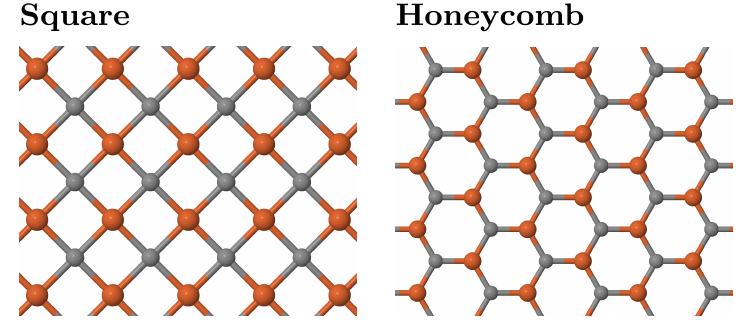}
	\caption{Iron carbides with square and honeycomb arrangements. The gray balls are C and the orange balls are Fe.}
	\label{fig:Fe_carbide_structures}
\end{figure}

Another possibility is that the experimentally observed structure is, in fact, an iron carbide. Zhao \textit{et al.} state that relatively small amounts of carbon may lie beyond the detection limits of their EELS setup and therefore cannot exclude the possibility that the membrane is made of iron carbide. It is also very difficult to observe C atoms near Fe in TEM due to the large difference in contrast. 
The iron carbides shown in Fig.~\ref{fig:Fe_carbide_structures} have binding energies per unit cell of -9.91 eV and -9.49 eV for the square and honeycomb lattice, respectively. The square lattice is thus the most stable configuration. The sum of the binding energy of separate monolayer Fe and graphene systems is -10.37 eV. The energy difference between the separate systems and the iron carbide is just 0.46~eV, which suggests that the iron carbide in square arrangement could be metastable. In particular, it is interesting to note that the lattice constant, i.e.\ the Fe-Fe distance, of the square iron carbide is 2.66 Å, which is extremely close to the experimentally observed value. However, since we find the structure to be, at best, metastable and no carbon signal was observed in EELS experiments, we are still skeptical that the observed structure is, in fact, iron carbide. 
More accurate measurements are needed in order to exclude the possibility of the membranes consisting of iron carbide.

\section{Embedded iron}
\begin{figure}[h]%
	\includegraphics{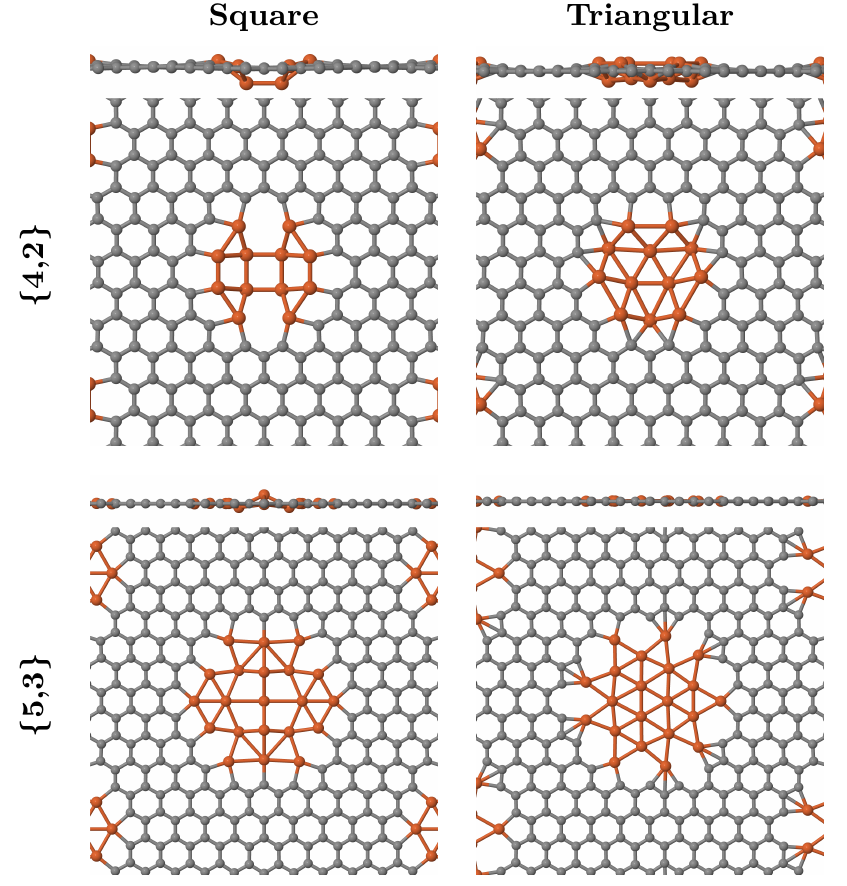}
	\caption{Top and side view of structurally relaxed graphene antidots with embedded Fe.}%
	\label{fig:FeGAL_output_structures}%
\end{figure}

We will now study the structural stability and magnetization of Fe membranes embedded in graphene perforations. In order to model this with DFT, we impose periodic boundary conditions, which means we effectively have a graphene antidot lattice (GAL), where the antidots are filled with Fe. 
We use the conventional $\{L,S\}$ notation to denote GALs with unit cell side length $L$ and antidot side length $S$, both in units of the graphene lattice constant, consistent with earlier work\cite{trolle2013large}. By filling a given antidot with the same amount of Fe atoms in the square and triangular configurations, we can make a direct comparison of the stability of the two systems by comparing their binding energies. In particular, we compare 12 and 21 Fe atoms embedded in a \{4,2\} and a \{5,3\} antidot lattice with hexagonal hole geometry, respectively. These antidot lattices are chosen because both square and triangular lattice configurations with an equal amount of Fe atoms can be found that conform fairly well with the antidots.  Figure~\ref{fig:FeGAL_output_structures} shows the structures after relaxation of all atoms in the unit cell.  
The figure shows that the surrounding graphene is almost unaffected by the presence of Fe, due to the large in-plane strength of graphene. It is also seen that the Fe bulges out-of-plane for the small antidots, especially for Fe in square arrangement. This indicates that the square lattice does not conform as well to the graphene lattice as the triangular lattice does for the small antidot. In the larger antidot, the Fe is seen to be mostly co-planer with the graphene, which indicates that both lattice configurations conform better to the graphene lattice. The Fe still bulges slightly out-of-plane in the square lattice configuration, which indicates that the square lattice still conforms worse to the graphene lattice than the triangular lattice. 
By comparing the binding energies of the two systems, we can determine which of the Fe configurations is more stable.

The unit cells we consider are probably too small for the spins to be decoupled between neighboring cells. This means that the magnitude of the magnetic moment may differ for isolated Fe membranes in graphene. However, due to the high strength of the supporting graphene lattice, we expect that structural properties will be in quantitative agreement with isolated Fe membranes.

We find that the triangular lattice is favored in the \{4,2\} antidot lattice with a binding energy difference of 2.31 eV, while the square lattice is favored in the \{5,3\} antidot lattice with a binding energy difference of 1.37 eV. The fact that the square lattice is favored in the large antidot, despite conforming worse to the graphene lattice, indicates that the square lattice has a larger binding energy to graphene than the triangular lattice. We therefore presume that the square lattice will have a greater advantage in larger antidots, where it conforms better to the graphene lattice. However, when the Fe membrane grows too large, the "bulk" behavior should overcome edge or interface effects, which should lead to formation of the triangular Fe lattice. Moreover, there is still the possibility that a 3D nanocrystal could form instead of the triangular monolayer membrane as the 3D structure, in principle, has lower energy than the 2D counterpart for sufficiently large structures. We thus speculate that there is an antidot size regime, where the square Fe lattice is favored, but when the antidots become too large, either the triangular monolayer Fe lattice or a 3D nanocrystal will be formed instead. 
However, we cannot investigate the extent of this regime further, due to the computational complexity of the DFT calculations.

\begin{figure}[htb]
	\centering
	\includegraphics{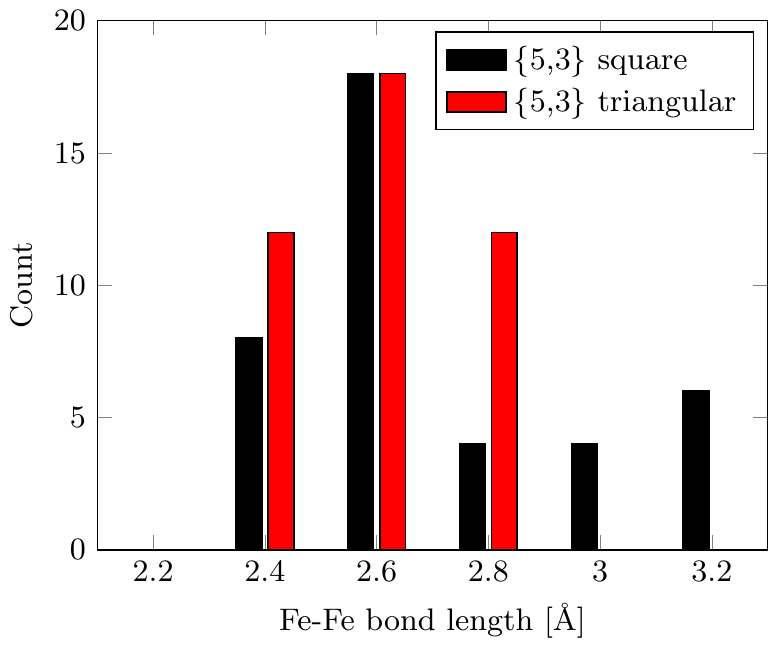}
	\caption{Fe-Fe bond lengths of the two \{5,3\} structures shown in Fig. \ref{fig:FeGAL_output_structures}.}
	\label{fig:bondlength}
\end{figure}

We saw previously that there was a rather large discrepancy between the bond lengths of the bulk monolayer Fe and the one measured in the experiments. To further investigate this discrepancy we have counted all the Fe-Fe bond lengths in the two \{5,3\} antidot structures in Fig.~\ref{fig:bondlength}. The figure shows that the Fe-Fe bond length inside the graphene antidots is generally quite close to the one measured experimentally, with a mean value of 2.7~Å and 2.6~Å in the square and triangular cases, respectively. The square lattice is thus strained by about 16\% on average compared to the bulk monolayer value. By comparison, the mean C-C bond length is almost unaffected by the interface with a mean value of 1.43~Å in both cases.

\begin{figure}[h]
	\centering
	\includegraphics{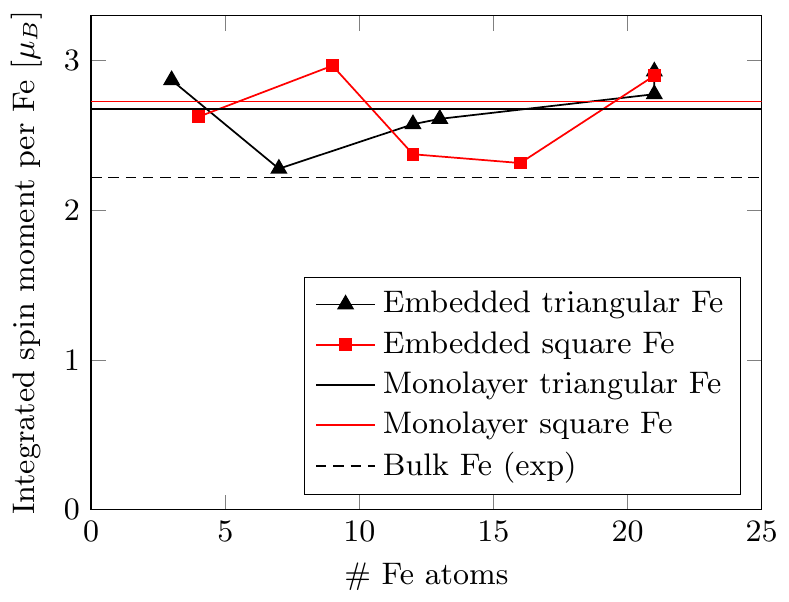}
	\caption{Integrated spin moment per atom for Fe membranes embedded in graphene antidots.}
	\label{fig:spin_per_fe}
\end{figure}

\begin{figure}[t]
	\centering
	\includegraphics{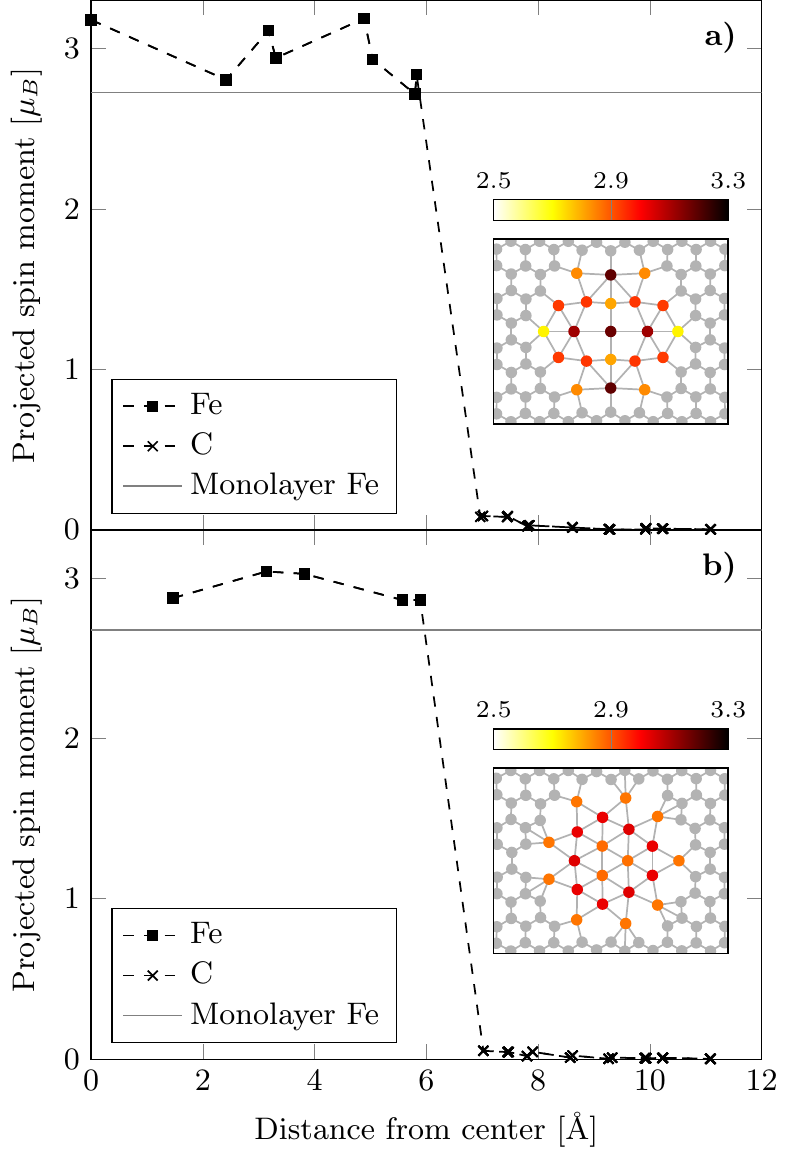}
	\caption{Projected spin moment for a \{5,3\} graphene antidot lattice with 21 Fe atoms in a hexagonal antidot in a) square arrangement and b) triangular arrangement.}
	\label{fig:L5_Fe21_projected_spin}
\end{figure}

Figure~\ref{fig:spin_per_fe} shows that the spin moment per Fe atom embedded in graphene antidots is around the value of monolayer Fe even for very few embedded Fe atoms. In contrast to Fe in a graphene monovacancy, where the spin moment is vanishing, the spin moment is only weakly affected by the presence of carbon on the edge. In fact, the spin moment may in some cases even exceed the monolayer value, due to the increased bond lengths. This is consistent with the result for Fe in a graphene divacancy, where the spin moment is also only weakly affected by the presence of carbon. This effect can be seen directly in Fig.~\ref{fig:L5_Fe21_projected_spin}, which shows the projected spin moment as a function of distance from the center of the antidot for a \{5,3\} graphene antidot lattice with 21 Fe atoms. The projected spin moment is calculated by integrating the difference in spin-up and spin-down electron densities inside the Voronoi volume associated with each atom. The figure shows that there is, in fact, an enhanced spin moment on nearly all Fe atoms in this case.

\section{Conclusions}
We have studied the stability of monolayer Fe and graphene-embedded Fe through \textit{ab initio} calculations. We find that the most stable configuration of monolayer Fe is the ferromagnetic triangular lattice with a lattice constant of 2.44~Å. This is in contrast to experimental results of graphene-embedded Fe, which shows that these structures have a square lattice configuration with a bond length of 2.65~Å. However, we find that the square lattice configuration has a lower edge formation energy. This means that, during formation, it might be favorable to form the square lattice and the structure could then be kinetically hindered from subsequently rearranging to the triangular lattice. Furthermore, we have compared the stability of the square and triangular Fe lattices in two different graphene antidot lattices. 
In the larger one of these, the square lattice is, in fact, more stable than the triangular lattice, with a mean Fe-Fe bond length of 2.7~Å. This result is in very close agreement with the experimental results. 
Our results show that only a few Fe atoms in the graphene antidots are sufficient to give rise to magnetic moments, which are comparable to the magnetic moment of monolayer Fe.

\section*{Acknowledgments}
The authors gratefully acknowledge the financial support from the Center for Nanostructured Graphene (Project No. DNRF58) financed by the Danish National Research Foundation and from the QUSCOPE project financed by the Villum Foundation.


\bibliography{../../literature}

\end{document}